\documentclass[submission,copyright,creativecommons]{eptcs}
\providecommand{\event}{PLACES 2010} 
\usepackage{breakurl}

\usepackage{color}
\usepackage{prooftree}
\usepackage{alltt}
\usepackage{multirow}
\usepackage{graphicx}
\def\midtilde{\raise-0.55ex\hbox{\textasciitilde}} 

\definecolor{dgreen}{rgb}{ 0, 0, 0} 

\newcommand{\fullss}{{\tt full-sessions} }

\newcommand{\la}{$\leftarrow$}

\newcommand{\st}[1]{{{\small {\tt #1}}}}

\newcommand{\new}{\mbox{\tt new}}
\newcommand{\unit}{\mbox{\tt ()}}
\newcommand{\send}{\mbox{\tt send}}
\newcommand{\recv}{\mbox{\tt recv}}
\newcommand{\sendS}{\mbox{\tt sendS}}
\newcommand{\recvS}{\mbox{\tt recvS}}

\newcommand{\inact}{\mbox{\tt inact}}
\newcommand{\sel}{\mbox{\tt sel}}
\newcommand{\selA}{\mbox{\tt sel1}}
\newcommand{\selB}{\mbox{\tt sel2}}
\newcommand{\offer}{\mbox{\tt offer}}

\usepackage{listings}
\lstnewenvironment{code}
    {\lstset{}%
      \csname lst@SetFirstLabel\endcsname}
    {\csname lst@SaveFirstLabel\endcsname}
\newtheorem{definition}{Definition}
\newtheorem{theorem}{Theorem}

\begin{document}

%
\title{Session Type Inference in Haskell}

\def\titlerunning{Session Type Inference in Haskell}

%
\author{Keigo Imai\\
      \institute{IT Planning Inc., Japan}
      \email{imai@nagoya-u.jp}
\and
Shoji Yuen \qquad\qquad Kiyoshi Agusa\\
      \institute{Graduate School of Information Science, Nagoya University, Japan}
      \email{yuen@is.nagoya-u.ac.jp\quad\qquad agusa@is.nagoya-u.ac.jp}
}

\def\authorrunning{Imai, Yuen, and Agusa}

\maketitle

%
\begin{abstract}
We present an inference system for a version of the $\pi$-calculus in Haskell for the session type proposed by Honda et al.
The session type is very useful in checking if the communications are well-behaved.
The full session type implementation in Haskell was first presented by Pucella and Tov,
which is `semi-automatic' in that the manual operations for the type representation was necessary.  
We give an automatic type inference for the session type by using a more abstract representation for the session type based on
the `de Bruijn levels'.  
We show an example of the session type inference for a simple SMTP client.
\end{abstract}

\section{Introduction}
The {\it Session-type system} \cite{honda_98_language} provides a way to statically check communication protocols.
Incorporating session types in the existing programming languages eases the communication centric programming in that session typed components are guaranteed to behave correctly by their types.
However, it is not apparent how to integrate the session typing discipline with the existing programming languages.

Several session-type implementations \cite{neubauer_04_implementation, pucella08session, sackman08session} have been proposed  for Haskell.
The {\it type-level programming} is shown to implement session types.
It is natural to use the functional dependencies \cite{fundeps} for encoding the duality of session types.
The indexed monad \cite{indexedmonad1,indexedmonad2} is used to propagate the session-type information through process constructs.

Currently, all existing im\-ple\-men\-ta\-tions \cite{neubauer_04_implementation, pucella08session, sackman08session} of the session type implementation require some manual annotation in program code to infer types.
The session types in \cite{neubauer_04_implementation} and \cite{sackman08session} often make even a simple program to be unnecessarily verbose.
The typechecking in \cite{pucella08session} requires incomprehensible annotation when the number of channel increases.
A fully-automatic type inference is essential as seen in other typed frameworks such as parser combinators \cite{parsec} and database access \cite{haskelldb}.

Our goal is to provide a fully-automatic session-type inference in Haskell.
We extend the work by Pucella and Tov \cite{pucella08session} to infer types without manual operations.
We show an implementation technique for the original session-type system \cite{honda_98_language} as the target language.

\paragraph{The issue of type-level representation}
The common idea in \cite{pucella08session} and \cite{sackman08session} is to track session types for multiple channels using the extra {\it symbol table} embedded in the Haskell's type.
The inferred Haskell type for a process would be \st{P} $\{c_1 \mapsto s_1; c_2 \mapsto s_2, \cdots\}$ where \st{P} is a process type constructor and $\{\cdots\}$ is the symbol table to assign each channel $c_i$ to its session type $s_i$.
This symbol table is represented in the {\it type-level}, hence the channels $c_i$ is not a value, but a {\it type} which reflects an identity of a channel.

In the implementation in \cite{pucella08session}, {\it type variables} represents channels in a symbol table.
To distinguish them from each other, such type variables are locally quantified at the position the channels are introduced.
Such a type variable is matched against the symbol table every time a type inference rule is applied.
Since the symbol table itself can only be represented as a type-level list of key-value pairs, matching on channels is unavoidable.
But in the Haskell type-level programming, such matching operation is not provided.

To alleviate this difficulty, \cite{pucella08session} devises the {\it stack manipulation} \st{dig} and \st{swap} on a symbol table for reordering.
The stack restricts the communication primitives only to the first entry of the symbol table.
\st{swap} swaps the first entry of the symbol table with the second one.
\st{dig} $p$ allows $p$ to access on the second entry of the symbol table.

In \cite{pucella08session}, it is stated that the automatic application of these operations is not possible without adding extra information in the symbol table and the answer for this problem is not shown.

\paragraph{Main idea}
To resolve the type matching problem, we use the natural number based on {\it de Bruijn level} as the type-level representation for channels.
The symbol table is represented as just the type-level list of session types, and accessed by the numbers.
As the number-based access on the type-level list is possible in the existing technique \cite{hlist}, type inference is fully automatic.


Our main contribution is to show an automatic inference of the session-type inference in Haskell.
\footnote{A working implementation, \fullss, which can be compiled by the Glasgow Haskell Compiler 6.10.2 or higher is available at:
\url{http://hackage.haskell.org/package/full-sessions/}. Typing {\tt cabal install full-sessions} in a shell will install \fullss in your environment.}
Although in \cite{pucella08session} only the capability passing is possible, our calculus is possible to pass a channel.  To show that our improvement is purely in the sense of matching, 
It is shown that by extending typing discipline in  \cite{pucella08session} it can have the same expressiveness as ours.



\paragraph{Related work}
Neubauer and Thiemann \cite{neubauer_04_implementation} implemented session types on a single communication channel in Haskell.
Their implementation avoids aliasing by prohibiting explicit use of a channel. 

Pucella and Tov \cite{pucella08session} have shown a general technique to encode session types in languages like Haskell, ML, and C\#.
Their implementation based on manual stack manipulations {\tt swap} and {\tt dig} liberates from type-level programming which is only available in Haskell, 
hence their technique can be applied to other languages which have parametric or generic types.
On the other hand, their implementation cannot enjoy fully-automatic type inference.

The implementation proposed by Sackman and Eisenbach \cite{sackman08session} supports full functionality of session types. 
However, their library requires a manual construction of session types.
There are trade-offs between such a manual handling and annotated type-inference approach in that 
while type-inference reduces unneeded annotations, explicit annotation with a rich set of syntax increases readability and expressiveness of types. 
We will discuss this aspect in the later section.

The difference of our implementation from the previous work is summarized in the following table.

\begin{center}
\begin{tabular}{l|c|c|c}
& channel &\multirow{2}*{annotation} & \multirow{2}*{portable} \\
& passing &   &  \\
\hline
Neubauer et al. \cite{neubauer_04_implementation}  & no & auto & no\\
Pucella et al. \cite{pucella08session}  & yes in a limited context\footnotemark & stack based channel handling & yes\\
Sackman et al. \cite{sackman08session} & yes & {manual construction of session types} & no\\
Our implementation & {\it yes}  & {\it auto} & {\it no}
\end{tabular}
\footnotetext{See section \ref{expressiveness}}
\end{center}

\paragraph{Paper Organization}
The rest of this paper is organized as follows. 
In Section \ref{sessionintro}, session types and the $\pi$-calculus is introduced from \cite{honda_98_language}.
In Section \ref{inference}, we show the session-type inference in Haskell, and compare it with other implementations.
We describe an example of a SMTP client using our implementation in Section \ref{network}.
In Section \ref{expressiveness}, we show that our implementation is more expressive than \cite{pucella08session} in the aspect of channel-passing
and show the way to extend \cite{pucella08session} to have the same expressive power as ours.
In Section \ref{discussion}, we discuss a few aspects of usability of session-type implementation.
Finally, Section \ref{concluding} concludes the paper.

\section{Session types and the $\pi$-calculus}
\label{sessionintro}
\subsection{The $\pi$-calculus}
The syntax of our $\pi$-calculus processes is defined by the following grammar:
\begin{eqnarray}
\begin{array}{rcl}
P ::= \ \send _\pi\ c\ e\ P\ |\ \recv _\pi\ c\ (\lambda x.P)\ |\ \selA _\pi\ c\ P\ |\ \selB _\pi\ c\ P\ \ |\ \offer _\pi\ c\ P_1\ P_2\\
  |\ \ \ \sendS _\pi\ c\ c'\ P\ |\ \recvS _\pi \ c\ (\lambda c'.P)\ |\ P\ |||\ Q\ |\ {\tt inact}_\pi\ |\ \new _\pi\ (\lambda c.P)\qquad\quad
\end{array}\nonumber
\end{eqnarray}
We use $\lambda$-abstraction to represent bindings using higher-order abstract syntax \cite{hoas}.
$x$ ranges over variables of basic values and channels, $c$ and $c'$ range over channels, and $P$ and $Q$ range over processes.
We put the subscript $\pi$ on each process constructor since they are overloaded in the later sections.

An input process ${\tt \recv _\pi}\ c\ {\tt (}\lambda x.P{\tt )}$ inputs a value via channel $c$, then binds it to $x$ in $P$. 
An output process {\tt $\send _\pi$ $c$ $e$ $P$} first evaluates $e$, then emits the value via channel $c$, and becomes $P$. 
{\tt sel1$_\pi$ $c$ $P$} and {\tt sel2$_\pi$ $c$ $P$} denote the selection of branch label $1$ or $2$ on $c$. 
It first sends the selected label, and becomes $P$. 
{\tt \offer$_\pi$ $c$ $P_1$ $P_2$} receives a label. 
Then it becomes $P_1$ or $P_2$, depending on the received label. 
{\tt \sendS $_\pi$ $c$ $c'$ $P$} sends channel $c'$ on $c$ and becomes $P$. 
{\tt \recvS $_\pi$ $c$ (}$\lambda c'.P${\tt )} receives a channel on $c$, binds it to $c'$ in $P$, and becomes $P$. 
These operations enable higher-order session communications.
$P\ |||\ Q$ runs $P$ and $Q$ concurrently. 
${\tt inact}_\pi$ is the constant to denote the terminated (inactive) process. 
{\tt new$_\pi$ (}$\lambda c.P${\tt )} generates a fresh channel $c$ bound in $P$. 

The operational semantics of the $\pi$-calculus is in Figure \ref{sos}. 
Here, $e\downarrow v$ represents that $e$ is evaluated to a value $v$.
The structural congruence of processes is in Figure \ref{strcong}.
The function fn$(P)$ denotes free names in $P$. 
$\equiv _\alpha$ denotes $\alpha$-equivalence.

\begin{figure}[tbh]
\begin{displaymath}
\mbox{\footnotesize {\sc Com}}:\ 
\begin{prooftree}
e \downarrow v
\justifies
{\tt \send _\pi}\ c\ e\ P\ |||\ {\tt \recv _\pi}\ c\ (\lambda x.Q) \longrightarrow P\ |||\ Q\{v/x\}
\end{prooftree}
\end{displaymath}
\begin{displaymath}
\mbox{\footnotesize {\sc Label}}_1:\ 
\begin{prooftree}
\justifies
{\tt sel1 _\pi}\ c\ P\ |||\ {\tt \offer _\pi}\ c\ Q_1\ Q_2 \longrightarrow P\ |||\ Q_1
\end{prooftree}
\end{displaymath}
\begin{displaymath}
\mbox{\footnotesize {\sc Label}}_2:\ 
\begin{prooftree}
\justifies
{\tt sel2 _\pi}\ c\ P\ |||\ {\tt \offer _\pi}\ c\ Q_1\ Q_2 \longrightarrow P\ |||\ Q_2
\end{prooftree}
\end{displaymath}
\begin{displaymath}
\mbox{\footnotesize {\sc Pass}}:\ 
\begin{prooftree}
\justifies
{\tt \sendS _\pi}\ c\ c'\ P\ |||\ {\tt \recvS _\pi}\ c\ (\lambda c'.Q) \longrightarrow P\ |||\ Q
\end{prooftree}
\end{displaymath}
\begin{displaymath}
\mbox{\footnotesize {\sc Scop}}:\ 
\begin{prooftree}
P \longrightarrow P'
\justifies
\mbox{{\tt new$_\pi$ (}$\lambda c.P${\tt )}} \longrightarrow \mbox{{\tt new$_\pi$ (}$\lambda c.P'${\tt )}}
\end{prooftree}
\hspace{3em}
\mbox{\footnotesize {\sc Par}}:\ 
\begin{prooftree}
P \longrightarrow P'
\justifies
P\ |||\ Q \longrightarrow P'\ |||\ Q
\end{prooftree}
\end{displaymath}
\begin{displaymath}
\mbox{\footnotesize {\sc Str}}:\ 
\begin{prooftree}
P \equiv P'\ \wedge\ P'\ \longrightarrow\ Q'\ \wedge\ Q'\equiv Q
\justifies
P\ \longrightarrow\ Q
\end{prooftree}
\end{displaymath}
\caption{The operational semantics of the $\pi$-calculus\label{sos}}
\end{figure}

\begin{figure}[tbh]
\begin{center}
$   P\equiv Q\ \mbox{if}\ P\equiv _\alpha Q \quad\quad    P\ |||\ {\tt inact} \equiv P \quad\quad P\ |||\ Q \equiv Q \ |||\  P \quad\  {\tt new _\pi}\ (\lambda c.{\tt inact}) \equiv {\tt inact}$\\
$   (P \ |||\  Q) \ |||\  R \equiv P \ |||\  (Q \ |||\  R) \qquad   {\tt new _\pi}\ (\lambda c.P) \ |||\  Q \equiv {\tt new _\pi}\ (\lambda c.P\ |||\ Q)\ \ \mbox{if } c \not\in\mbox{fn}(Q) $
\caption{Structural congruence of the $\pi$-calculus processes\label{strcong}}
\end{center}
\end{figure}

\subsection{Session types}
\label{sec:sessiontypes}
In this subsection and following subsection, we review a session type system in \cite{honda_98_language}.

A session type represents a protocol which is associated with an endpoint of a channel. 
$v$ ranges over types for basic values, and $u$ ranges over session types.
The session types in this paper are defined by the following grammar:
\begin{eqnarray}
\begin{array}{rcl}
u &::= &{\tt Send }\ v\ u\ |\ {\tt Recv}\ v\ u\ |\ {\tt Select}\ u_1\ u_2\ |\ {\tt Offer}\ u_1\ u_2\\
&| & {\tt Throw }\ u_1\ u_2\ |\ {\tt Catch}\ u_1\ u_2\ |\ {\tt End}\ |\ {\tt Bot}
\end{array}\nonumber
\end{eqnarray}
{\tt Send $v$ $u$} denotes a protocol to emit a value of type $v$ followed by a behavior of type $u$.
{\tt Recv $v$ $u$} denotes a protocol of receiving a value of type $v$ followed by a behavior of type $u$.
${\tt Select}\ u_1\ u_2$ denotes to be either behavior of type $u_1$ or type $u_2$ after emitting a corresponding label $1$ or $2$.
${\tt Offer}\ u_1\ u_2$ denotes a behavior like either $u_1$ or $u_2$ according to the incoming label.
{\tt Throw $u_1$ $u_2$} denotes a behavior to output of a channel with session type $u_1$ followed by a behavior of type $u_2$. 
{\tt Catch $u_1$ $u_2$} is the input of a channel with session type $u_1$ followed by a behavior of type $u_2$. 
{\tt End} denotes a terminated session. 
{\tt Bot} is the type for a channel whose endpoints are already engaged by two processes, so that no further processes can own that channel. 
For example,  in $({\tt \send _\pi}\ c \ e\ {\tt inact}\ |||\ {\tt \recv _\pi}\ c\ (\lambda x.{\tt inact}))$, $c$ has the session type {\tt Bot}. 

A session type $u$ has the dual $\overline{u}$. 
The definition of dual is illustrated in Figure \ref{duality}.
A dual of a session on one end of a channel is the session on the other end of the same channel.
\begin{figure}[tbh]
\begin{eqnarray}
\begin{array}{rclrclrcl}
\overline{{\tt Send}\ v\ u} &=& {\tt Recv}\ v\ \overline{u} &
 \overline{{\tt Select}\ u_1\ u_2} &=& {\tt Offer}\ \overline{u_1}\ \overline{u_2}  &
 \overline{{\tt Throw}\ u_1\ u_2} &=& {\tt Catch}\ u_1 \overline{u_2}\\
\overline{{\tt Recv}\ v\ u} &=& {\tt Send}\ v\ \overline{u} &
 \overline{{\tt Offer}\ u_1\ u_2 } &=& {\tt Select}\ \overline{u_1}\ \overline{u_2} &
 \overline{{\tt Catch}\ u_1\ u_2 } &=& {\tt Throw}\ u_1\ \overline{u_2}\\
\overline{\tt End} &=& {\tt End} & \overline{\tt Bot}&=&{\it undefined}&  &
\end{array}\nonumber
\end{eqnarray}
\caption{Duality for session types \label{duality}}
\end{figure}

\subsection{The typing rules}
In the session-type system \cite{honda_98_language}, there are two kinds of type judgments, value judgment  $\Gamma \vdash e \triangleright v$ and process judgment $\Gamma \vdash P \triangleright \Delta$.  A process  $P$ is {\it typeable} if there exists some $\Gamma, \Delta$ such that $\Gamma \vdash P \triangleright \Delta$.
$\Gamma$ denotes  {\it sorting} that maps variables to types of basic values. 
$\Delta$ denotes {\it session type environment} or {\it session typing} that maps names to session types. 
A {\it completed} type environment is the one that assigns the type {\tt End} to every name appearing in a process.

A process is {\it typeable} under $\Gamma$, iff $\Gamma \vdash P \triangleright \Delta$ for a given $\Delta$. 
A typeable process never fails due to communication mismatch.

The typing rules are defined in Figure \ref{typingrules}. 
The composition of type environments  $\Delta \oplus \Delta'$ is defined by the component-wise extension of the type algebra which is defined as follows:

\begin{eqnarray}
\begin{array}{cc}
{\tt End} \oplus u = u & u \oplus {\tt End} = u\\
u \oplus u' = {\tt Bot} & \mbox{if}\ \overline{u}=u' \ \ \mbox{otherwise}\ {\it undefined}
\end{array}\nonumber
\end{eqnarray}

The literature \cite{honda_98_language} defines an erroneous  process using following notions: A {\it $c$-process} is a process prefixed by subject $c$, such as ${\tt \send _\pi}\ c\ e\ P$ and ${\tt \recvS _\pi}\ c\ (\lambda c'.P)$.
A {\it $c$-redex} is the parallel composition of two $c$-processes in either of form ${\tt \send _\pi}\ c\ e\ P\  |||\ {\tt \recv _\pi}\ c\ (\lambda x.Q)$, 
${\tt sel}_i\ c\ P\ |||\ {\tt \offer _\pi}\ c\ Q_1\ Q_2$ for $i\in\{1,2\}$, or ${\tt \sendS _\pi}\ c\ c'\ P\ |||\ {\tt \recvS _\pi}\ c\ (\lambda c'.Q)$.

\begin{definition}[Error]
We shall say that $P$ is an {\it error} if $P \equiv {\tt new _\pi}\ (\lambda \tilde{c}. Q\ |\ R)$ where $Q$ is, for some $c$, the parallel composition of 
either two $c$-processes that do not form a $c$-redex, or three or more $c$-processes.
\end{definition}

Then we quote the following theorem, which is also valid for our framework, from \cite{honda_98_language}
\footnote{To show this we do not require type preservation, as stated in \cite{giunti09session}.}:
\begin{theorem}[Type Safety]
A typeable program never reduces to an error.
\end{theorem}

\begin{figure}[tbh]
\begin{displaymath}
\begin{prooftree}
\Gamma \vdash e \triangleright v \quad\quad\quad \Gamma \vdash  P \triangleright \Delta \cdot c:u
\justifies
\Gamma \vdash {\tt \send _\pi}\ c\ e\ P \triangleright \Delta \cdot c:\ \verb!Send!\ v\ u
\using
\mbox{\sc [Send]}
\end{prooftree}
\quad\quad
\begin{prooftree}
\Gamma, x:v \vdash  P \triangleright \Delta \cdot c:u
\justifies
\Gamma \vdash {\tt \recv _\pi}\ c\ (\lambda x.P) \triangleright \Delta \cdot c:{\tt Recv}\ v\ u
\using
\mbox{\sc [Rcv]}
\end{prooftree}
\end{displaymath}
\begin{displaymath}
\begin{prooftree}
\Gamma \vdash P \triangleright \Delta \cdot c:u_1
\justifies
\Gamma \vdash {\tt sel1 _\pi}\ c\ P \triangleright \Delta \cdot c:{\tt Select}\ u_1\ u_2
\using
\mbox{\sc [Sel]}
\end{prooftree}
\quad\quad
\begin{prooftree}
\Gamma \vdash P \triangleright \Delta \cdot c:u_1 \quad\quad\quad \Gamma \vdash Q \triangleright \Delta \cdot c:u_2
\justifies
\Gamma \vdash {\tt \offer _\pi}\ c\ P\ Q\triangleright \Delta \cdot c:{\tt Offer}\ u_1\ u_2
\using
\mbox{\sc [Br]}
\end{prooftree}
\end{displaymath}
\begin{displaymath}
\begin{prooftree}
\Gamma \vdash P \triangleright \Delta \cdot c:u_2
\justifies
\Gamma \vdash {\tt \sendS _\pi}\ c\ c'\ P \triangleright \Delta \cdot c:{\tt Throw}\ u_1\ u_2 \cdot c' : u_1
\using
\mbox{\sc [Thr]}
\end{prooftree}
\end{displaymath}
\begin{displaymath}
\begin{prooftree}
\Gamma \vdash P \triangleright \Delta \cdot c:u_2, c':u_1
\justifies
\Gamma \vdash {\tt \recvS _\pi}\ c\ (\lambda c'. P)\triangleright \Delta \cdot c:{\tt Catch}\ u_1\ u_2
\using
\mbox{\sc [Cat]}
\end{prooftree}
\end{displaymath}
\begin{displaymath}
\begin{prooftree}
\Gamma \vdash P \triangleright \Delta \quad\quad\quad \Gamma \vdash Q \triangleright \Delta'
\justifies
\Gamma \vdash P\ |||\ Q \triangleright \Delta \oplus \Delta'
\using
\mbox{\sc [Conc]}
\end{prooftree}
\qquad
\begin{prooftree}
\Gamma \vdash P \triangleright \Delta \cdot c:{\tt Bot}
\justifies
\Gamma \vdash {\tt new _\pi}\ (\lambda c. P)\triangleright \Delta
\using
\mbox{\sc [Cres]}
\end{prooftree}
\qquad
\begin{prooftree}
\Delta\ \mbox{completed}
\justifies
\Gamma \vdash {\tt inact}\triangleright \Delta
\using
\mbox{\sc [Inact]}
\end{prooftree}
\end{displaymath}
\caption{Typing rules for session types\label{typingrules}}
\end{figure}

\section{Session-type inference on Haskell}
\label{inference}
We first introduce concurrency primitives in our implementation using the $\pi$-calculs defined in the Section 2.1.
Then we present a few techniques to embed session types in Haskell as in \cite{neubauer_04_implementation} and \cite{pucella08session}.
Finally we show the session type reconstruction for multiple channels based on de Bruijn levels. 

\subsection{Concurrency primitives and session types in \fullss}
Our implementation, \fullss, provides concurrency primitives using {\it monad} rather than the syntax provided in the Section 2.
This is because monad is the most well-known way to describe communicating processes in Haskell.

To keep connection between the original session-type system with our implementation, we show our primitives using continuation monad.
The behaviour of each primitives is captured by the continuation-passing monad of type \st{((a -> Pi d$_1$) -> Pi d$_2$)} where \st{Pi d$_i$} corresponds to the type of process term $P$ in Section 2.1, and \st{d$_i$} is a type-level representation of a session-type environment $\Delta$.
The meaning of each primitives are summarized in the Table \ref{primitives}.
In the table we abuse the $\lambda$-notation of hoas syntax in Section 2 to represent a syntactic function from values or channels to processes.
$k$ ranges over continuations of type \st{a -> Pi d1}.
For readability, we use the \st{ixdo} notation \cite{pucella08session}, which provides a syntactic sugar to write programs in an imperative style.
For example,  the term \st{ixdo send $c$ $e$; recv $c$} and \st{ixdo fork (send $c$ $e$); recv $c$} are interpreted as $\lambda k.\send_\pi\ c\ e\ (\recv_\pi\ c\ k)$ and $\lambda k.(\recv _\pi\ c\ k\ |||\ \send _\pi\ c\ e)$, respectively.

Processes can be run using the function \st{runS}. \st{runS $p$} runs a $\pi$-calculus process $p (\lambda \_. \inact _\pi)$.
Hereafter we call the all primitives in Table \ref{primitives} as a {\it session} of type \st{Session}.

\begin{table}[htb]
{\small
\begin{center}
\begin{tabular}{r|l|l}
& Function & Meaning\\
\hline
{\it Channel Creation} & {\tt new} & $\lambda k. \new _\pi\ k$\\
{\it Value Output} & {\tt send $c$ $e$}  & $\lambda k. \send _\pi\ c\ e\ (k\ \unit)$ \\
{\it Value Input} &{\tt recv $c$}   & $\lambda k. \recv _\pi\ c\ k$\\
{\it Selection} & {\tt sel$i$ $c$}\ \ $(i\in\{\mbox{\tt 1},\mbox{\tt 2}\})$  & $\lambda k. \sel i_\pi\ c\ (k\ \unit)$ \\
{\it Offering} & {\tt offer $c$ $p_1$ $p_2$}   & $\lambda k.\offer _\pi\ c\ (p_1\ k)\ (p_2\ k)$ \\
{\it Session Delegation} & {\tt sendS $c$ $c'$}  & $\lambda k. \sendS _\pi\ c\ c'\ (k\ \unit)$\\
{\it Session Reception} &{\tt recvS $c$}   & $\lambda k.\recvS _\pi\ c\ k$\\
{\it Fork} & {\tt fork $p$} & $\lambda k.((k\ \unit)\ |||\ (p\ (\lambda \_.\inact _\pi)))$\\
{\it Calling Haskell I/O} & {\tt io $m$} & (Execute Haskell's {\tt IO} action $m$ and pass the result to the continuation) \\
{\it Recursion of a session} & {\tt recur1 $f$ $c$} & $\lambda k.f\ c\ k$\ \ (Recursive call of a session $(f\ c)$ where $c$ is a channel)\\
\multirow{2}{*}{\it Recursive use of a channel} & \multirow{2}{*}{\tt unwind $c$} & $\lambda k. k\ \unit$(Unwind a recursive session type \\
& & {\tt Rec $n$ $u$} into $u$[{\tt Var $n$} $\mapsto$ {\tt Rec $n$ $u$}] on $c$
\end{tabular} 
\caption{Primitives in the \fullss library\label{primitives}}
\vspace{-1em}
\end{center}
}
\end{table}

\subsection{Session type inference for a single channel}
\subsubsection{A single-threaded participant}
\label{evolution}
Let us begin with a case of single channel in a single-threaded participant.
In such a case a session type {\it advances} as a session proceeds. For example a type \st{Send Int End} advances to \st{End} when a channel of that type is used to send an integer.
To track such an advance of a session type, we assign a pair of session types, called a {\it pre-type} and a {\it post-type}, to each occurrence of a channel. 
A pre-type denotes the session type {\it before} a session starts.
Similarly, a post-type is the session type {\it after} a session ends.

In many cases post-types act as a {\it placeholder}, which allows concatenation of two session types.
For example, consider one of the simplest sessions, \st{send $c$ True}.
The pre-type of the channel $c$ in this session is \st{Send Bool $u$} and the post-type of it is $u$, where $u$ is a type variable.
This means that another session which uses the channel $c$ can be further concatenated after this session.

The concatenation of two session types are done by unification. In a concatenation \st{$s_1$;$s_2$} of two sessions, the post-types of channels in $s_1$ is unified with the pre-types of ones in $s_2$.
The pre-types of channels in the concatenated session \st{$s_1$;$s_2$} is same as the ones in \st{$s_1$}. The post-types of channels in \st{$s_1$;$s_2$} is the ones of \st{$s_2$}.
Accordingly, \st{(send $c$ True; send $c$ "abc")}
has \st{(Send Bool (Send String $u_2$))} as the pre-type
and $u_2$ as the post-type on the channel \st{$c$}, where $u_2$ is a type variable distinct from $u$.

For a more complex example, the code below describes a simple calculator server. 
\begin{code}
  server c = ixdo x <- recv c; y <- recv c; offer c (send c (x+y::Int)) (send c (x<y))
\end{code}
The \st{server} firstly receives two values of type \st{Int} and a branch label (here the label is either {\sf 1} or {\sf 2}),
then sends an answer either of type \st{Int} or of \st{Bool} according to the label.

The pre/post-type of the channel \st{c} in the \st{server} can be inferred by the GHC's typechecker via auxiliary function \st{channeltype1}. 
By showing the type of \st{(channeltype1 server)} using GHC's interactive environment, users will obtain 
the following response:
\begin{code}
  prompt> :t channeltype1 server
  channeltype1 server :: (Recv Int (Recv Int (Offer (Send Int a) (Send Bool a))), a)
\end{code}

%

\subsubsection{Duality of two session types}
\label{duality}
The \st{fork} primitive requires the {\it duality} between pre-types of two sessions.
Here we explain it by using the previous example of a calculator server.
Firstly, a client of the server would be like this:
\begin{code}
  client c = ixdo send c 123; send c 456; sel2 c; ans <- recv c; 
                  io (putStrLn (if ans then "Lesser" else "Greater or Equal"))
\end{code}
The pre-/post-type of \st{c} in \st{client} is \st{(Send Int (Send Int (Select $u_1$ (Recv Bool $u$)))} and $u$, respectively. By putting \st{server} and \st{client} in parallel by \st{fork}, and by generating a channel by \st{new}, we obtain the code below:
\begin{code}
  calc c    = ixdo fork (server c); client c;  
  startCalc = ixdo c <- new; calc c
\end{code}
The above code typechecks because the two usages of \st{c} in \st{client} and \st{server} are dual.
The resulting pre-type is \st{Bot}, as the session-type algebra of \cite{honda_98_language} implies. 
The post-type is \st{End} since \st{fork} requires the usage of channels in the given session to be ended.
\footnote{Such discipline can also be observed in session-type systems equipped with a thread-spawning construct,
the ``ended" condition of {\bf Spawn} rule in \cite{coppo07asynchronous}.}
Here we confirm it:
\begin{code}
  prompt> :t channeltype1 calc
  channeltype1 calc :: (Bot, End)
\end{code}
A session can be run by the function \st{runS}. Typing {\tt runS startCalc} will produce the result ``\st{Lesser}" on the console.
The following is the result of the execution using the interpreter:
\begin{code}
  prompt> runS startCalc
  Lesser
\end{code}

\subsection{Tracking sessions with multiple channels by De Bruijn indexing}
\label{multiple}
\begin{figure}[thb]
\begin{center}
\includegraphics[width=7cm,keepaspectratio]{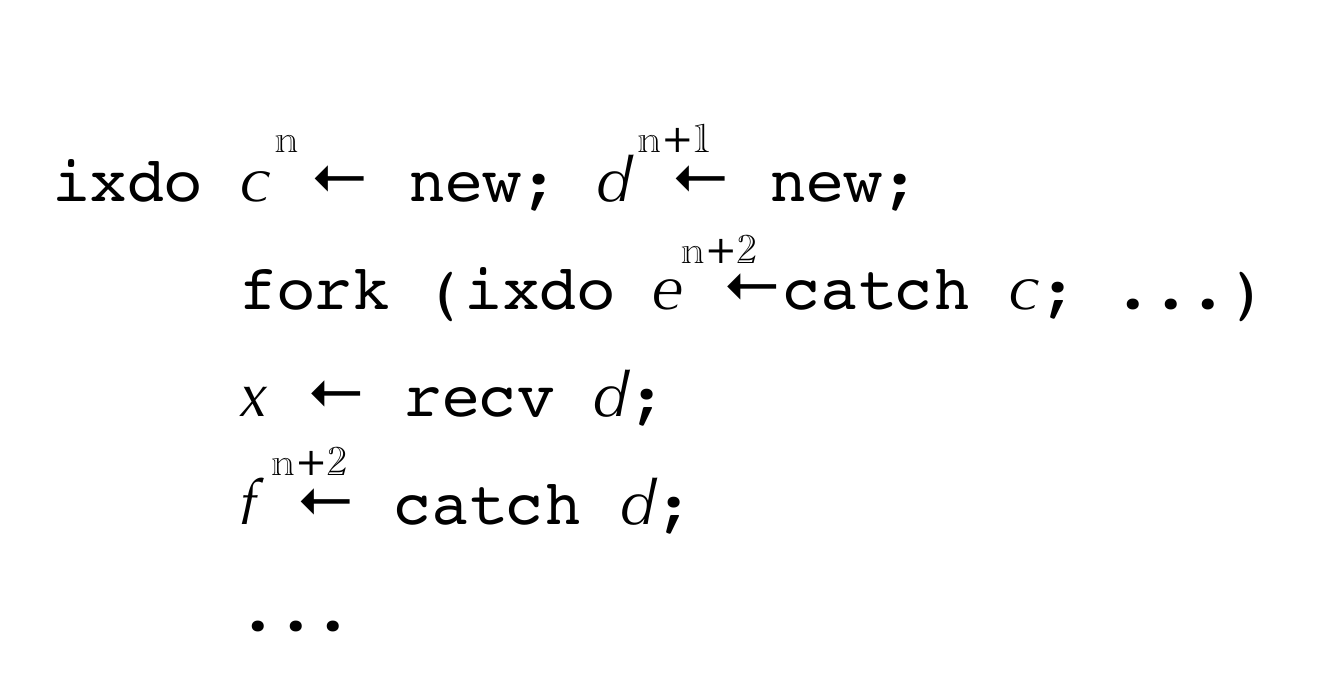}
\caption{De Bruijn {\it level} indexing in a session\label{pic:debruijn}}
\end{center}
\end{figure}

To track usages of multiple channels in type-level, a natural number of {\it de Bruijn level} is assigned to each channel.
De Bruijn level represents the nesting depth of a variable binder. 
For example, in a $\lambda$-calculus term $\lambda x.\lambda y.x$ the level of the variable $x$ is $0$ whereas $y$ is $1$.
Figure \ref{pic:debruijn} shows the de Bruijn level indexing of a session.
In the figure, the de Bruijn level of a variable is denoted by a superscript at the binding position.
Note that we need to count on only channels, 
hence each variable $c, d, e$ and $f$ have an index but $x$ does not. 

De Bruijn levels are assigned to the type of channels.
A channel has the type of the form \st{Channel $t$ $n$} where $n$ is a de Bruijn level of the channel and 
$t$ is a ``type-tag" \cite{kiselyov08lightweight}. 
We do not explain the type-tag, since it is out of scope of our paper.
Natural numbers are represented by combinations of the two types representing peano-numerals \st{Z} and \st{S $n$} where each of them denotes $0$ and $n+1$ respectively. 
For example, a channel which has de Bruijn level $2$ has type \st{Channel $t$ (S (S Z))}.
Each number points to a certain position of a type environment. 

Session types of multiple channels are recorded in extra type environments.
We need two type environments for pre-types and post-types.
Hereafter we call them {\it pre-environment} and {\it post-environment}, respectively.

\begin{figure}[thb]
\begin{center}
\includegraphics[width=12cm,keepaspectratio]{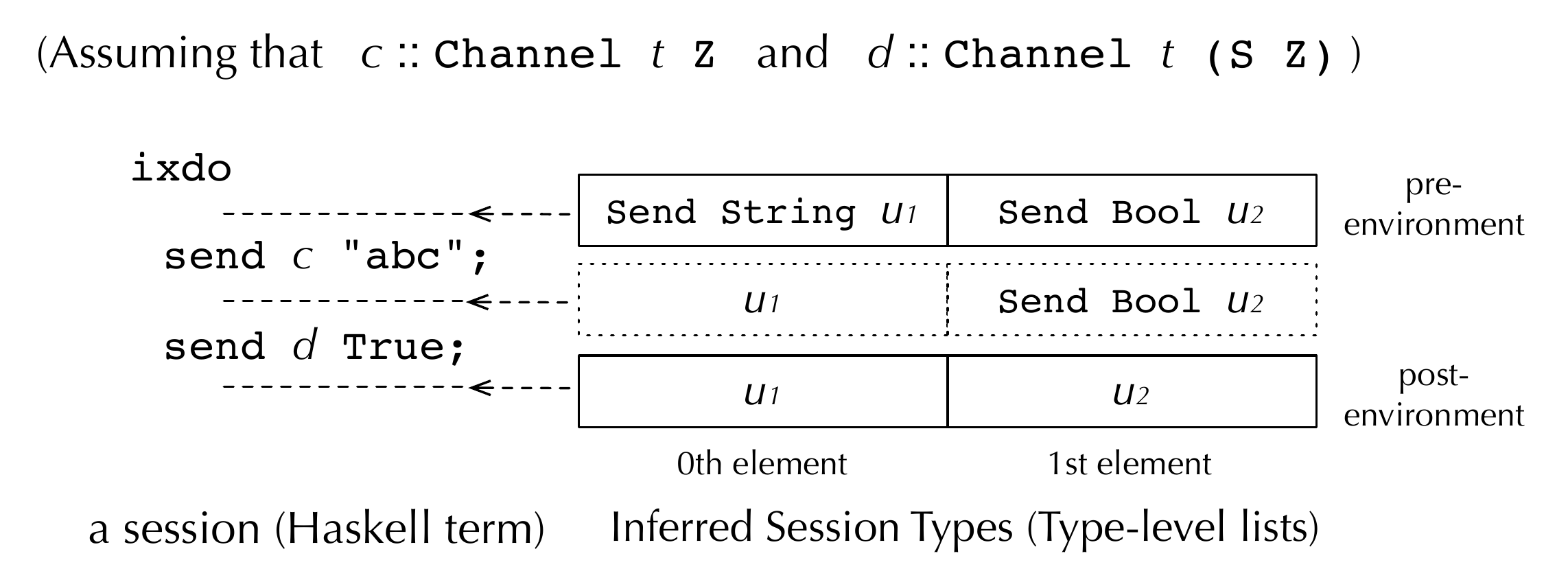}
\caption{Tracking session types by numbers\label{pic:multichan2}}
\end{center}
\end{figure}

Such an environment is represented by a list of session types, and its elements are accessed by specifying the number of de Bruijn level.
Figure \ref{pic:multichan2} is an example of a session \st{send $c$ "abc"; send $d$ True} and its pre-/post-environment.
Assuming that $c$ and $d$ have (Haskell-) type \st{Channel $t$ Z} and \st{Channel $t$ (S Z)} respectively, 
the pre- and post-environment of the session is inferred as shown in the figure.
$c$ and $d$ has pre-type \st{Send String $u_1$} and \st{Send Bool $u_2$}, 
and post-types of them are \st{$u_1$} and \st{$u_2$}, respectively.
Note that the figure also depicts the session-types in an intermediate step after \st{send $c$ "abc"}.
In that state, $c$ has type \st{$u_1$} and $d$ has type \st{Send Bool $u_2$}.


The type of a session has the form of \st{Session} $t$ $ss$ $tt$ $a$.
The pre-/post-environments are at the position of $ss$ and $tt$ respectively.
The parameter $a$ is the type of a value returned by a session, and $t$ is a type-tag. 

The pre-/post-environments $ss$ and $tt$ are actually represented by {\it type-level list}s \cite{hlist}.
A type-level list is either \st{$ss$ :> $u$} or \st{Nil}, where $ss$ is another type-level list and $u$ is a session type, and \st{Nil} is a empty list. 
Note that the type constructor \st{:>} is left-associative, for example \st{$ss$:>$a$:>$b$} is interpreted as \st{($ss$:>$a$):>$b$}. 
Also note that the type environment is counted from left to right order. 
For example, the 0-th element of \st{Nil:>$a$:>$b$:>$c$} is $a$. 

Provided that the type of $c$ is \st{Channel $t$ Z} and the type of $d$ is \st{Channel $t$ (S Z)}, a session of the previous example \st{(send $c$ "abc"; send $d$ True)} has type \st{Session $t$ (Nil:>Send String $u_1$:> Send Bool $u_2$) (Nil:>$u_1$:>$u_2$) ()}.
  
In general, the de Bruijn levels can be a non-constant value, like \st{$n+1$}, \st{$n+2$} and so on.
For example, if the length of a session-type environment \st{$ss$} is \st{$n$}, and the type of $c$ and $d$ is \st{Channel $t$ $n$} and \st{Channel $t$ (S $n$)} respectively, a session \st{(send $c$ 1; send $d$ True)} has type \st{Session $t$ ($ss$:>Send Int $u_1$:> Send Bool $u_2$) ($ss$:>$u_1$:>$u_2$) ()}. 
Constraints for the length of a session-type environment is represented in the type-level by the type-class \st{SList $ss$ $n$}, which represents that the length of \st{$ss$} is \st{$n$}.
Observe that the existence of the placeholder \st{$ss$} in each of session-type environments. 
This makes possible to handle arbitrary numbers of channels by concatenation of sessions which introduce new channels, which involves unification between the post-environment of the earlier session and the pre-environment of the later session.

When a new channel is introduced, post-environments are {\it extended} to store the session type of the introduced channel. 
The primitive \st{new} and \st{catch} involve such a mechanism. 
\st{new} has pre-environment \st{$ss$} and post-environment \st{$ss$:>Bot.}
At the same time \st{new} returns a channel of type \st{Channel $t$ $n$}, where $n$ is equal to the length of $ss$ and points to the leftmost position of the post-environment, namely \st{Bot}.
Hence the index of a generated name is assured to be fresh.

\begin{figure}[thb]
\begin{center}
\includegraphics[width=14cm,keepaspectratio]{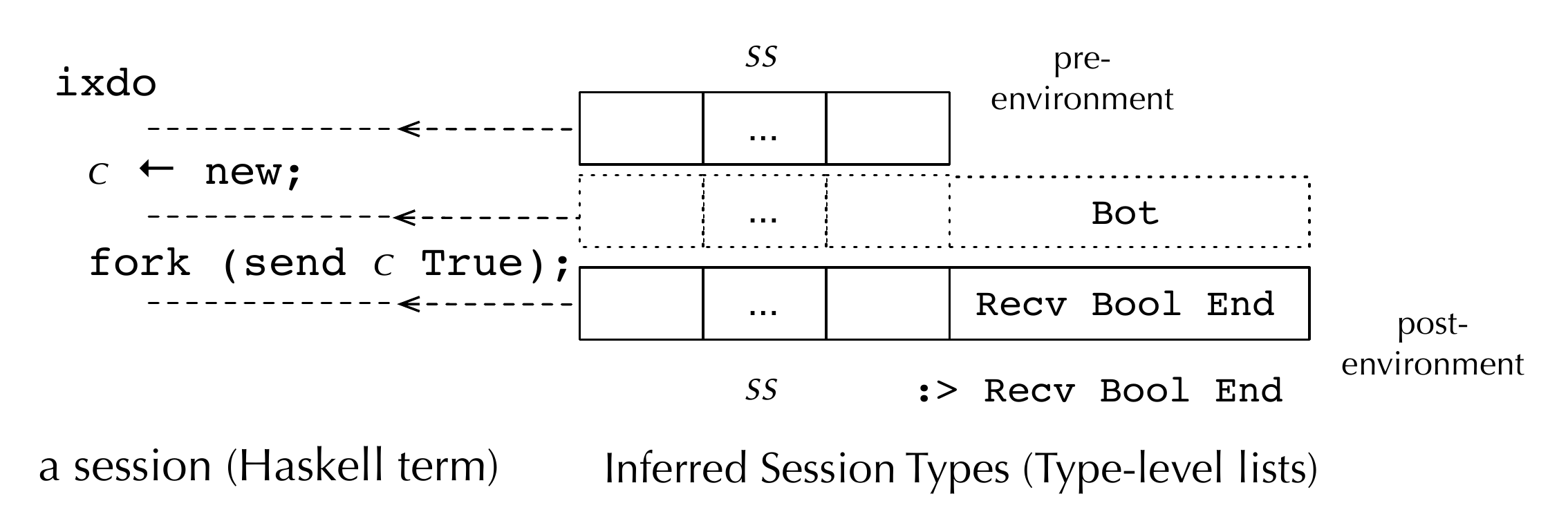}
\caption{Extension of a type environment by \st{new} operation\label{pic:multichan4}}
\end{center}
\end{figure}

Figure \ref{pic:multichan4} shows the pre-/post-environments of a session \st{($c$ \la\ new; fork (send $c$ True))}.
The post-environment has an extra entry for the newly created channel.
The post-type of the newly created channel is dual of \st{Send Bool End}, which is required to communicate with the forked session.

\begin{figure}[thb]
\begin{center}
\includegraphics[width=16cm,keepaspectratio]{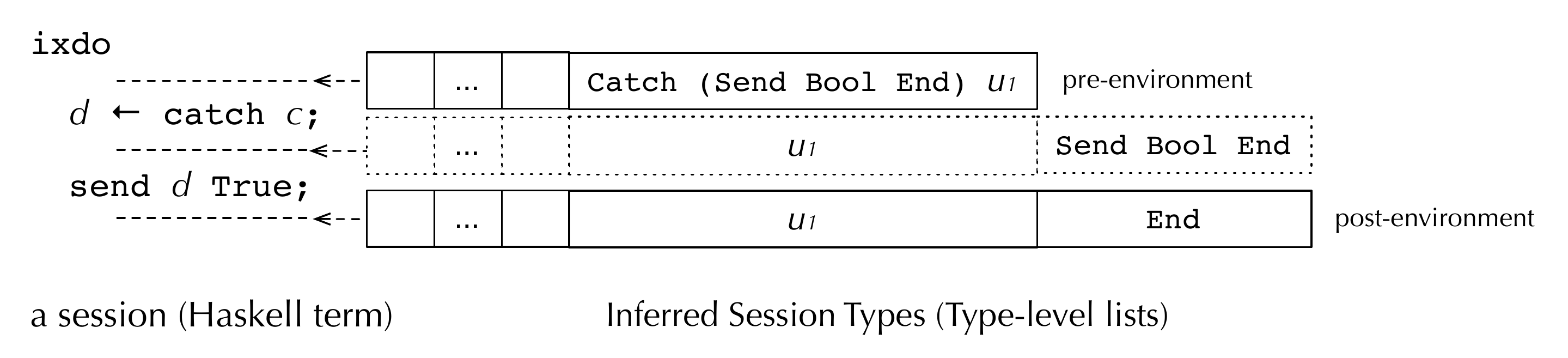}
\caption{Extension of a type environment by \st{catch} operation\label{pic:multichan5}}
\end{center}
\end{figure}

Similarly, \st{catch $c$} has the pre-/post-environment \st{$ss$} and \st{$tt$:>$u'$}, where $n$-th element of \st{$ss$} is \st{Catch $u'$ $u$} and that of \st{$tt$} is $u$.
Figure \ref{pic:multichan5} shows such use of \st{catch} and the inferred session types.


%



\subsection{Comparison of existing Haskell implementations of session types}
\label{comparison}

Our encoding based on de Bruijn indexing reduces most of annotations which are required in the other works.
We show that by giving a few examples of sessions.

\paragraph{Stack-based implementation}
The implementation by Pucella and Tov \cite{pucella08session} applies a stack of session types as the representation of a type-environment.
Communication primitives can only access the top of the stack, hence explicit manipulation of stack is required.
The combinator \st{dig} and \st{swap} is provided for such purpose.
The \st{swap} combinator swaps the top two channels on the stack.
On the other hand, \st{dig} combinator converts a given session to operate on a deeper channel stack.
Provided that the session type for $c$ and $d$ is on the top of the stack in this order, 
the code below is equivalent to \st{(send c "abc"; send d True)}:
\begin{code}
  ixdo send c "abc"; swap; send d True
\end{code}
or
\begin{code}
  ixdo send c "abc"; dig (send d True)
\end{code}
As a number of channels increases, more stack operations will be required.
In \cite{pucella08session} a few approaches to this problem are discussed, 
however the problem had been left open, and our number-based approach is not covered. 

\paragraph{Manual construction of session types}
The implementation by Sackman and Eisenbach \cite{sackman08session} provides a very rich set of communication primitives,
at a cost of manual construction of session types.
There are two communication media, channels and {\it Pid}s.
We show the simplest case of communication via Pid.
The example below passes an integer $10$ to the other thread and terminates. 
It is equivalent to \st{runS (ixdo c <- new; fork (send c 10); recv c)}.
\begin{code}
  (s, a) = makeSessionType (
             newLabel ~>>= \a -> 
             a .= send (undefined :: Int) ~>> end 
             ~>> sreturn a)
  
  p = run s a (ssend 10) srecv
\end{code}
Here \st{makeSessionType} returns a collection of session types $s$ and its fragment $a$.
In the argument of \st{makeSessionType} the construction of a session type is described procedurally.
Again, as a number of threads with different protocol increases, the more construction of session types will be required.
The case of channel-based communication is similar.

\section{An example SMTP client}
\label{network}
We show the network functionality of the \fullss by the example of a SMTP client with multiple channels. 
A single-channel version of SMTP client with session types has its origin from \cite{neubauer_04_implementation}. 

Table \ref{nwprimitives} shows additional primitives for network communication. 
To model network protocols, the {\it type-based branching}, \st{sel$i$N} and \st{offerN}, is provided in addition to the previous {\it label-based branching}.
Note that the \st{sel$i$N} does nothing, but we need them to infer the session types for type-based selections.

\begin{table}[thb]
{\small
\begin{center}
\begin{tabular}{r|l|l}
& Syntax & Meaning\\
\hline
Connect to a service  & {\tt $c$ $\leftarrow$ connectNw $s$} & Connect to a service $s$ and bind a session channel to $c$\\
Type-based offering  & {\tt offerN $c$ $p_1$ $p_2$} & Offer two receiving session $p_1$ and $p_2$ on $c$ \\
\multirow{2}{*}{Selection annotation} & \multirow{2}{*}{\tt sel$i$N $c$\ \ $(i\in\{\mbox{\tt 1},\mbox{\tt 2}\})$} & Determine which branch of {\tt Select $u_1$ $u_2$} will be\\
& &  selected on $c$
\end{tabular} 
\caption{Additional primitives for network programming\label{nwprimitives}}
\end{center}
}
\end{table}

Here we show our implementation of SMTP client in the simplest form. 
Firstly, the types for SMTP commands and replies are defined as follows:
\begin{code}
  -- Types for SMTP commands. 
  newtype EHLO = EHLO String
  newtype MAIL = MAIL String 
  newtype RCPT = RCPT String
  data    DATA = DATA        
  data    QUIT = QUIT        
  newtype MailBody = MailBody [String]
  
  -- Types for SMTP server replies (200 OK, 500 error and 354 start mail input)
  newtype R2yz = R2yz String; newtype R5yz = R5yz String; newtype R354 = R354 String
\end{code}
To deal with the stream-based communication of TCP, either a parser or a printer for each type of communicated values must be prepared.
Provided such functions exist, the SMTP client is described as follows:
\begin{code}
  -- auxiliary functions
  send_receive_200 c mes = ixdo send c mes; (R2yz _) <- recv c; ireturn ()
  send_receive_354 c mes = ixdo send c mes; (R354 _) <- recv c; ireturn ()
  
  sendMail c d = ixdo  -- the body of our SMTP client
    (R2yz _) <- recv c                   -- receive 220
    send_receive_200 c (EHLO "mydomain") -- send EHLO, then receive 250
    unwind0 c                            -- (annotation) unwind a recursion variable
    sel1N c                              -- (annotation) branch to send 'MAIL FROM'
    from <- recv d                          -- (1) input the sender's address on d
    send_receive_200 c (MAIL from)       -- send 'MAIL FROM', then receive 250
    unwind1 c                            -- (annotation) unwind a recursion variable
    sel1N c                              -- (annotation) branch to send 'RCPT TO'
    to <- recv d                            -- (2) input the recipient's address on d
    send c (RCPT to)                     -- send 'RCPT TO'
    offerN c (ixdo                       -- branch the session according to the reply
      (R2yz _) <- recv c                 -- if 250 OK is offered
      sel1 d; mail <- recv d                -- (3) input the content of the mail on d 
      unwind1 c                          -- (annotation) unwind a recursion variable
      sel2N c                            -- (annotation) branch to send 'DATA'
      send_receive_354 c DATA            -- send 'DATA' and receive 354
      send_receive_200 c (MailBody mail) -- send the content of the mail
      unwind0 c                          -- (annotation) unwind a recursion variable
      sel2N c                            -- (annotation) branch to send 'QUIT'
      send c QUIT; close c               -- send 'QUIT' and close the connection
     ) (ixdo 
      (R5yz errmsg) <- recv c;           -- if 500 ERROR is offered
      sel2 d; send d errmsg;                -- (4) output the error message on d
      send c QUIT; close c)              -- send 'QUIT' and close the connection
    close d
\end{code}
The \st{sendMail} takes two channels \st{c} and \st{d} as its parameters. The former is used to communicate with the SMTP server while latter is used to prepare necessary information for sending a mail.
By checking the type of \st{typecheck2 sendMail}, the following type is answered by GHC:
\begin{code}
  typecheck2 sendMail :: (SList ss l, IsEnded ss b1) => Session t
    (ss :> Recv R2yz (Send EHLO (Recv R2yz (Rec Z (SelectN 
            (Send MAIL (Recv R2yz (Rec (S Z) (SelectN 
              (Send RCPT (OfferN (Recv R2yz (Var (S Z))) (Recv R5yz (Send QUIT Close))))
              (Send DATA (Recv R354 (Send MailBody (Recv R2yz (Var Z)))))))))
            (Send QUIT Close)))))
        :> Recv String (Recv String (Select (Recv [String] Close) (Send [String] Close))))
    (ss :> End :> End) ()
\end{code}
The SMTP protocol is successfully represented in the pre-type of \st{c}. 
A server that have the dual of this type can communicate with this client.

Observe that the two channels are used with no annotation.
On the other hand, the implementation of \cite{pucella08session} requires the \st{swap} operation before and after the each occurrence of \st{d}, namely at \st{(1)}, \st{(2)}, \st{(3)} and \st{(4)}, and if we add more channels, more complicated bookkeeping operations will be required.


\section{Expressiveness of the encoding based on de Bruijn levels\label{expressiveness}}
We discuss the expressiveness between our implementation and the others.
The discussion goes around the feature of higher-order sessions originally provided in \cite{honda_98_language}.
We show that the presentation provided in \cite{pucella08session} has some limitation.
Due to that fact, our implementation is more expressive than \cite{pucella08session}.

At the same time, we sketch that their \st{swap} and \st{dig} can provide the same expressive power as our de Bruijn based solution.


\subsection{The limitation of capability-passing primitive}
The implementation presented in \cite{pucella08session} does not provide primitives for channel-passing,
while they provide \st{send\_cap} and \st{recv\_cap} which communicate the {\it capability} of channels.
We discuss here that capability-passing does not provide full-fledged feature of the higher-order session.

The primitives \st{send\_cap} and \st{recv\_cap} only synchronize on a given channel, but not communicate any run-time information.
Instead, on the synchronization the sender's side delegates the capability of a channel to the receiver's side.

In several cases this capability-passing is succinct to simulate name-passing.
Here we sketch their capability-passing primitives by rewriting the code in our implementation using \st{send\_cap} and \st{recv\_cap}.
The following code in our implementation
\begin{code}
  pq c = ixdo fork (q c); p c
  p c  = ixdo d <- new; throw c d; str <- recv d; io (putStrLn str)
  q c  = ixdo d <- catch c; send d "Hello"
\end{code}
will be rewritten in their framework as follows\footnote{Assume they provide \st{new} and \st{fork} primitive in their language.}:
\begin{code}
  pq' c  = ixdo d <- new; fork (q' c d); p' c d
  p' c d = ixdo send_cap c; dig (ixdo str <- recv d; io (putStrLn str))
  q' c d = ixdo recv_cap c; dig (send d "Hello")
\end{code}
Notice that we put the channel-creation primitive (\st{new}) {\it before} the forking.

The problem of the capability passing is that the communicated channel must be known {\it before} the run-time.
This is fatal for the distributed application which can not communicate any information before run-time.
In addition, the $\pi$-calculus theory says that under the existence of recursion (or replication), the rewriting shown above is  not possible.
See the following code.
The process sends fresh channel on \st{c} repeatedly with sending "Hello" on the thrown channel.
\begin{code}
  loop c = ixdo unwind c; d <- new; throw c d; send d "Hello"; recur1 loop c
\end{code}
The \st{new} is now put under the loop. 
Such repeated channel-creation cannot bring out of the loop, hence cannot be expressed in \cite{pucella08session}.

\subsection{Implementing channel-passing with \st{dig} and \st{swap}}
The problem of \st{send\_cap} is that the type-signature of it requires {\it static} (type-level) identity of channels, which seems to be unneeded constraint with respect to the original Session-type system \cite{honda_98_language}.
The following is the type signature for the both primitives.
\begin{code}
  send_cap :: Channel t -> Session (Cap t e (Cap t' e' r' :!: r), (Cap t' e' r', x)) 
                                   (Cap t e r, x) 
                                   ()
  recv_cap :: Channel t -> Session (Cap t e (Cap t' e' r' :?: r), x)
                                   (Cap t e r, (Cap t' e' r', x)) 
                                   ()
\end{code}
Here, the first arguments of them must have pre-type \st{Cap t e (Cap t' e' r' :!: r)} or \st{Cap t e (Cap t' e' r' :?: r)}, where \st{t'} denotes identity of communicated channel.

Since there is no reason to have it, we put the alternative capability type \st{Cap2 e' r'} which does not have identity of a channel.
By replacing \st{Cap} with \st{Cap2}, we get the following signature for sender's side:
\begin{code}
  send_chan :: Channel t -> Channel t' 
                         -> Session (Cap t e (Cap2 e' r' :!: r), (Cap t' e' r', x)) 
                                    (Cap t e r, x) 
                                    ()
\end{code}

On the other hand, the receiver's side is rather complicated. 
Since the identity of the communicated channel was lost at the sender's side, we must give the new one on the receiver's side.
That was done by introducing an universally quantifying type variable on the signature.
Since the type variable cannot escape from its scope, the continuation of the process must be given as the second argument.
\begin{code}
  recv_chan :: Channel t -> (forall t'. Channel t' -> 
                              Session (Cap t e r, (Cap t' e' r', x))
                                      (Cap t e rr, y) 
                                      () )
                         -> Session (Cap t e (Cap2 e' r' :?: r), x)
                                    (Cap t e rr, y)
                                    ()
\end{code}

\section{Other aspects of Session-type implementation}
\label{otheraspects}
\subsection{Representing recursion of session types}
Many literature on session types takes {\it equi-recursive} view of types \cite{TAPL}, which identifies $\mu t.T$ with its unfolded form $T\{\mu t.T/t\}$.
Unfortunately, Haskell and many other languages do not support such typing.
Hence ours and the other implementations of session types \cite{neubauer_04_implementation, sackman08session, pucella08session} take different approach, iso-recursive view of recursion on types.
In this subsection we review each of them.

\paragraph{The first implementation of recursive session types \cite{neubauer_04_implementation}}
Neubauer et al. invented a representation of recursive type which requires a new type declaration for each session-type recursion.
The type {\tt Rec} is a fixpoint type constructor defined as follows:
\begin{code}
  newtype Rec f = MkRec (f (Rec f))
\end{code}
{\tt Rec} has kind {\tt (* $\rightarrow$ *) $\rightarrow$ *}.
When a session repeatedly sends integers, the type corresponding to it must be declared first:
\begin{code}
  newtype G self = G (Send Int self)
\end{code}
where the type variable {\tt self} is a placeholder for the recursion variable.
By applying {\tt Rec} on this type, the type {\tt Rec G} is isomorphic to $\mu t.\mbox{\tt Send Int $t$}$.

Such a type is unwound by declaring type classes.
Consider expanding {\tt Rec G} to {\tt Send Int (Rec G)}.
A type class {\tt RECBODY} is declared as follows:
\begin{code}
  class RECBODY t c | t -> c where ...
\end{code}
The first type parameter {\tt t} is for folded form of a recursive type and the second type parameter {\tt c} is for unfolded form.
A functional dependency \cite{fundeps} {\tt t $\rightarrow$ c} declares that Haskell's type checker can automatically infer {\tt c} from {\tt t}.
The expanded form as an instance of {\tt RECBODY} becomes following:
\begin{code}
  instance RECBODY (Rec G) (Send Int (Rec G)) where ...
\end{code}

However, declaring such types and instances for each recursion seems redundant.
As \cite{neubauer_04_implementation} requires another explicit declaration of session types, such redundancy should be avoided.
Our implementation and the other two implementations do not require such extra declarations.
Hereafter we review that of ours and Pucella et al., though \cite{sackman08session} do not give any account of it.

\paragraph{Expansion of recursive type representations by type-level computation}
In our implementation, such a recursive session type is represented in the form of {\tt Rec Z (Send Int (Var Z))}. 
In {\tt Rec {\it n} {\it r}} a type parameter {\it n} denotes the de Bruijn level of the binder and {\tt Var {\it n}} is its occurrence.
Then {\tt Rec Z (Send Int (Var Z))} denotes $\mu t. \mbox{\tt Send Int }t$.
The primitive {\tt unwind} expands the recursion.
For example, {\tt Rec Z (Send Int (Var Z))} is expanded as {\tt Send Int (Rec Z (Send Int (Var Z)))}.
The other primitive {\tt recur1 {\it f} {\it c}} is behaviourally equal to {\it f c}, yet this ensures that the pre-types of the all channels other than {\it c} is ended.
Such annotation is required since sometimes the usage of some channels has no explicit end and it cannot be inferred by the typechecker.

Our encoding depends on Haskell's type-level computation.
The encoding by Pucella and Tov \cite{pucella08session} is not limited to Haskell, at a cost of a bit complex representation in types.

\paragraph{Deferred expansion of recursion body \cite{pucella08session}}
In \cite{pucella08session}, a recursive type is represented by using de Bruijn {\it indices} (not levels) as a binder for recursion variable.
Thus, the type for a session which repeatedly sends integers is {\tt Rec (Send Int (Var Z))} (note that this {\tt Rec} is different from previous one).
Here {\tt Var {\it n}} is a type-level recursion variable, where {\it n} is a peano-numeral of the de Bruijn index of the binder.

Their {\it capability type} has the form of {\tt Cap t e r} where {\tt t} is type tag \cite{kiselyov08lightweight} and {\tt r} is a session type.
The second parameter {\tt e} represents a {\it stack} which is used for bookkeeping during recursion.
Note that this stack is different from that of multiple channels in Section \ref{comparison}.

A recursive type is not immediately expanded, but deferred by using a notational trick.
To see this, consider expanding a recursive session {\tt Cap t e (Rec (Send Int (Var Z)))}.
The recursion body is put at the second parameter, resulting  {\tt Cap t (Send Int (Var Z),e) (Send Int (Var Z)))}.
When one met the recursion variable {\tt Cap t (Send Int (Var Z),e) (Var Z)}, the substitution is actually done and it becomes {\tt Cap t (Send Int (Var Z),e) (Send Int (Var Z))}.

By putting a notational trick, Pucella and Tov succeed to represent session-type recursion in a language-independent way.
Since our implementation already uses heavy type-level computation, we have used full functionality of type-classes to represent recursions in a more direct way.
In other words, our encoding does not require stacks for recursions.

\subsection{Inter-process Communication}
One notable difference between ours and \cite{sackman08session} is that the communication primitives in their implementation are based on process identities, {\it Pid}s..
Providing both Pids and channels as communication media would be much convenient in view of scalability.
However, since their framework requires much of annotations, usage of channels would become burden.

In \cite{sackman08session}, a typical inter-process communication example that a process {\tt parent} forks a process {\tt child} to send an integer {\tt 52} is written as follows:
\begin{code}
  (st, a) = makeSessionType (
              newLabel ~>>= \a ->
              a .= recv int ~>> send bool ~>> end ~>>
              sreturn a)
    where
      int = undefined :: Int
      bool = undefined :: Bool
  
  parent = fork a dual (cons (a, notDual) nil) child
           ~>>= \(_, childPid) ->
           createSession a dual childPid
           ~>>= \childCh ->
           withChannel childCh (ssend 52 ~>> srecv)
           ~>>= sliftIO . print
  
  child _ parentPid
         = createSession a notDual parentPid
           ~>>= \parentCh ->
           withChannel parentCh
             (srecv ~>>= ssend . ((==) 42))
\end{code}
where {\tt a} is the value-level session type associated to the channel.
{\tt st} is the session type associated to a Pid, which is not used.
The values {\tt dual} and {\tt notDual} is needed because the framework does not infer which side of the protocol it uses.
Actual communication is described at the second argument of each call of {\tt withChannel}.

This complication arises in order to maintain type-level symbol tables.
Comparing with this, the same behaviour can simply be described in our framework as follows:
\begin{code}
  parent = ixdo
    ch <- new;
    fork (child ch)
    send ch 52 >>> recv ch >>>= io . print
  
  child ch = recv ch >>>= send ch . ((==) 42)
\end{code}
Thanks to inference of the symbol table based on de Bruijn levels, our encoding requires essential communication primitives around the session type.

\section{Usability}
\label{discussion}
Here we discuss a few aspects of session type implementation.

\paragraph{Trade-offs between type inference and manual construction of session types}
As we have shown in Section \ref{comparison},  annotations required by our implementation is not more than any of the other implementations.
However, there seems to be a few  advantages in \cite{sackman08session} in a few points.
(1) Recursion of a session type is treated more naturally in \cite{sackman08session}. By using term-level operation for constructing session types, 
\cite{sackman08session} offers more readable formulation of recursion via labels.
As you can observe in the SMTP example of the previous section, recursion on a session type require a few of not so intuitive annotations \st{unwind$_i$} on the term-level to represent a recursive protocol.
(2) Manual construction of session types in term-level offers chance of subtyping. 
It is difficult to allow subtyping of session types in the parallel composition, 
because of our bijective encoding of duality to extract more information in a parallel composition of a session.

\paragraph{Readability of type error messages}
If the duality check of two session types fails, the type error would be reported.
For example, by replacing the occurrence of an integer \st{456} in Section \ref{duality} with a string \st{"456"}, the following error is obtained
:\footnote{Here, \st{[Char]} is a type synonym of \st{String}.}
\begin{alltt}
  examples/calc.hs:<xx>:0:
      Couldn't match expected type `[Char]' against inferred type `Int'
        Expected type: tt' :> Send [Char] a
        Inferred type: tt' :> Send Int (Select (Recv Int End) (Recv Bool End))
      When generalising the type(s) for `plus'
\end{alltt}
The error reports that the inferred pre-type of \st{client} is not compatible with the expected one. The position \st{<xx>} of the reported error is not at \st{send c "456"} itself, but at the position where the dual of the session type is calculated, namely the occurrence of the \st{fork}.
Thus, this error message directly shows which session types are not compatible. 
Even the type-level hackery we depend tends to produce large type signatures,  the type error itself can be concisely represented.

\section{Concluding remarks}
\label{concluding}
This paper showed a Haskell implementation of the session-type inference. 
Our implementation infers session types fully automatic without any manual operations such as stack operations in \cite{pucella08session}.



The treatment of binders is the key issue for embedding one language into another, as stated in \cite{poplmark}.
In our implementation, we took a separated approach for the {\it term-level computation} and {\it type-level} (compile-time) computation.
In the term-level computation, a fresh channel is represented by $\lambda$-abstraction (the technique usually called Higher order abstract syntax),
utilizing the power of variable-bindings in the host language of Haskell.
In the type-level computation, {\it de Bruijn levels} represent in channel types to compare names. These are the key to automate the session-type inference.
However, since the current technique depends on the type-level programming functionality of Haskell, it is not easy to export this technique to the other programming languages yet.

Our technique using de Bruijn level can be applied to other substructural type systems for the $\pi$-calculus, such as linear type systems and multiparty session types \cite{honda08multiparty}.
In particular, encoding of multiparty session types is promising.
The end-point session types of \cite{honda08multiparty} is much similar with the original binary session types \cite{honda_98_language}, hence our technique can be effectively used.
In Haskell, types cannot have different concrete representation of types since Haskell's type inference goes through unification.
However, due to asynchronous nature of multiparty session types, a end-point type $k\langle U\rangle;k'\langle U'\rangle;T$ can have different concrete representation 
$k'\langle U'\rangle;k\langle U\rangle;T$ if $k\not=k'$ where two first components can be exchanged.
To express such type in a unique form, again our de Bruijn encoding of channels might play a key role.
That is, ordering such asynchronous sequencing by the de Bruijn level, one can obtain the unique representation of a end-point type.
Yet much remains to be done in making such ideas in a real code.

{\bf Acknowledgments} 
The first author thanks his colleagues in IT Planning Inc. for supporting him during the writing of this paper.
This work was partially supported by the Grant-in-Aid (Scientific Research (B) 20300009 and  Scientific Research(C) 19500026) from the Ministry of Education, Culture, Sports, Science, and Technology of Japan.

%
\label{sect:bib}
\bibliographystyle{eptcs}
\bibliography{imai-places2010}

\end{document}